\documentclass[conference]{IEEEtran}
\IEEEoverridecommandlockouts
\usepackage{cite}
\usepackage{amsmath,amssymb,amsfonts}
\usepackage{algorithmic}
\usepackage{graphicx}
\usepackage{textcomp}
\usepackage{xcolor}
\usepackage{url}
\def\BibTeX{{\rm B\kern-.05em{\sc i\kern-.025em b}\kern-.08em
    T\kern-.1667em\lower.7ex\hbox{E}\kern-.125emX}}
\begin{document}

\title{A Zigbee Based Cost-Effective Home Monitoring System Using WSN}

\makeatletter
\newcommand{\linebreakand}{%
  \end{@IEEEauthorhalign}
  \hfill\mbox{}\par
  \mbox{}\hfill\begin{@IEEEauthorhalign}
}
\makeatother








\author{
    \IEEEauthorblockN{
        Garapati Venkata Krishna Rayalu\IEEEauthorrefmark{1}\IEEEauthorrefmark{2}, 
        Paleti Nikhil Chowdary\IEEEauthorrefmark{1}\IEEEauthorrefmark{3}, 
        Manish Nadella\IEEEauthorrefmark{1}\IEEEauthorrefmark{4},  \\
        Dabbara Harsha\IEEEauthorrefmark{1}\IEEEauthorrefmark{5}, 
        Pingali Sathvika\IEEEauthorrefmark{1}\IEEEauthorrefmark{6},
        B.Ganga Gowri\IEEEauthorrefmark{1}\IEEEauthorrefmark{7}
    }
    \IEEEauthorblockA{\IEEEauthorrefmark{1}Amrita School Of Artificial Intelligence, Coimbatore, Amrita Vishwa Vidyapeetham, India}
    \IEEEauthorblockA{
        Email: \IEEEauthorrefmark{7}gangab.90@gmail.com, \IEEEauthorrefmark{2}gkrishnarayalu@gmail.com, \IEEEauthorrefmark{5}dabbaraharsha@gmail.com,\\
        \IEEEauthorrefmark{3}nikhil.28@outlook.in, \IEEEauthorrefmark{4}manishnadella03@gmail.com, \IEEEauthorrefmark{6}pingalisathvika@gmail.com
    }
}


\maketitle

\begin{abstract}
WSNs are vital in a variety of applications, including environmental monitoring, industrial process control, and healthcare. WSNs are a network of spatially scattered and dedicated sensors that monitor and record the physical conditions of the environment.Significant obstacles to WSN efficiency include the restricted power and processing capabilities of individual sensor nodes and the issues with remote and inaccessible deployment sites. By maximising power utilisation, enhancing network effectiveness, and ensuring adaptability and durability through dispersed and decentralised operation, this study suggests a comprehensive approach to dealing with these challenges. The suggested methodology involves data compression, aggregation, and energy-efficient protocol. Using these techniques, WSN lifetimes can be increased and overall performance can be improved. In this study we also provide methods to collect data generated by several nodes in the WSN and store it in a remote cloud such that it can be processed and analyzed whenever it is required.

\end{abstract}


\section{Introduction}

Wireless Sensor Networks (WSNs) are networks of inexpensive, low-power, small devices with wireless communication, sensors, and microcontrollers. According to \cite{ref6} these networks have various applications, including environmental monitoring, industrial process control, and healthcare. WSNs play a crucial role in wildlife conservation by detecting and tracking animal movements, monitoring crop health in agriculture \cite{ref7}, and ensuring the stability of buildings and bridges \cite{ref8}. Additionally, WSNs can assist in home monitoring and support elderly individuals.

Small, battery-operated sensor nodes form the basis of Wireless Sensor Networks (WSNs), which use wireless communication to monitor and collect data from their surroundings. These sensor nodes continuously sense environmental conditions and gather data. The nodes form a network with different topologies, such as star, mesh, or cluster-tree, and communicate with one another via wireless protocols.
Energy-efficient communication protocols such as Zigbee and LoWPANs, along with techniques like data compression, have been used to improve network efficiency \cite{ref5}. The distributed and decentralized nature of WSNs, enabling local decision-making based on gathered data, enhances adaptability and robustness.

Considering the existing research, studies have focused on developing and enhancing energy-efficient protocols for WSNs \cite{ref6}. Other research has explored techniques such as data compression, aggregation, and hierarchical routing to improve overall network effectiveness \cite{ref8}. The impact of distributed and decentralized operation on adaptability and robustness has also been investigated \cite{ref5}. However, there is a lack of a comprehensive strategy that simultaneously addresses power and computational constraints, remote deployment challenges, energy efficiency, and decentralized operation. Therefore, the Zigbee protocol is chosen as a solution.

Objectives of our work is to put to use an energy-efficient protocol that optimizes computing resources and power usage in WSNs.Address challenges associated with remote and inaccessible deployment sites.Improve network efficiency through data aggregation. Design a distributed and decentralized strategy to enhance the robustness and adaptability of WSNs.

To achieve these goals, we propose a unique methodology that combines multiple approaches and readily available algorithms specifically tailored for WSNs. This methodology takes into account factors such as remote deployment sites, limitations of individual sensor nodes, and the need for energy-efficient operations. It leverages the benefits of distributed and decentralized decision-making to enhance adaptability and resilience.

The following sections in this paper section 2 outlines the setup, utilizing XBee modules and Arduino-integrated sensors for data collection. In Section 3, we develop a user-friendly GUI for efficient data visualization. Section 4 focuses on collecting and securely storing the data on the cloud, while implementing continuous monitoring for detecting data changes.

\section{Related Works}

The table of literature review Table \ref{tab:lit_rev} provides a summary of the key findings from the relevant research literature.

\begin{table*}[h!]
  \centering
  \caption{Literature Review}
  \label{tab:lit_rev}
  \begin{tabular}{|c|p{3cm}|p{3cm}|p{3cm}|p{3cm}|}
    \hline
    Ref no & Title & Author & Methodology & Results\\
    \hline
    \cite{ref1} & Xbee-Based WSN Architecture for Monitoring of Banana Ripening Process Using Knowledge-Level Artificial Intelligent Technique & S. Altaf, S. Ahmad, M. Zaindin, M.W. Soomro & In order to track banana ripening, the article suggests an Xbee wireless sensor network (WSN) design. In order to improve analysis and decision-making in this process, artificial intelligence techniques are used. However, it is devoid of precise methodological information. & According to the findings, a wireless sensor network based on Xbees and knowledge-level AI were able to accurately classify bananas into three states: normal, rotten, and unknown.\\
    \hline
    \cite{ref2} & Elderly Infrared Body Temperature Telemonitoring System with XBee Wireless Protocol & T. H. Y. Ling, L. J. Wong & The research describes an XBee wireless protocol-based elderly infrared body temperature telemonitoring system. It makes it possible to track and remotely monitor an elderly person's body temperature. The citation, however, is devoid of precise methodological information. & The study used an XBee wireless protocol-based telemonitoring system to successfully monitor elderly people's body temperatures.\\
    \hline
    \cite{ref3} & Ambient Assisted Living Environment Towards Internet of Things Using Multifarious Sensors Integrated with XBee Platform & Nagender Suryadevara, Sean Kelly, S.C. Mukhopadhyay & The study focuses on exploiting IoT technology to enhance the quality of life in assisted living settings by using the XBee platform and sensors to build an ambient assisted living environment. But no precise methodological information was offered. &  Using the XBee platform and many sensors, the study successfully developed an ambient assisted living environment. Through better monitoring and assistance, this integration increased the quality of life for residents of assisted living facilities.\\
    \hline
    \cite{ref4} & A Smart Monitoring System for Campus Using Zigbee Wireless Sensor Networks & Alaa Allahham, Md Arafatur Rahman  & The study describes a smart campus monitoring system based on Zigbee. It keeps track of temperature, humidity, light output, and sound levels using wireless sensor networks. The citation did not, however, provide any information regarding the process. & The study was successful in putting in place a smart campus monitoring system based on Zigbee. It makes real-time environmental parameter monitoring and data collecting possible, facilitating effective campus infrastructure management.\\
    \hline
    \cite{ref5} & Performance Analysis of Data Transmission on a Wireless Sensor Network Using the XBee Pro Series 2B RF Module & I Gusti Made Ngurah Desnanjaya et al. & The XBee Pro Series 2B RF module used in the article is used to analyze data transmission on a wireless sensor network. It assesses this module's effectiveness and dependability without presenting a thorough methodology or analysis. & The design and execution of wireless sensor networks are influenced by the examination of the XBee Pro Series 2B RF module's data transmission capabilities.\\
    \hline
    
    \end{tabular}
\end{table*}

\hfill

In \cite{ref1} Altaf et al. describes a wireless monitoring system for banana ripening using knowledge-level artificial intelligence algorithms and XBee-based WSN architecture. The system incorporates XBee modules for wireless communication between sensor nodes and a central controller, providing real-time data collecting. Throughout the ripening process, a variety of sensors are used to measure the temperature, humidity, and gas concentration levels. In order to determine the ideal ripening conditions, the collected data is processed using knowledge-level artificial intelligence techniques that incorporate professional knowledge and rules. The proposed system offers an efficient and optimized approach to monitor and control the banana ripening process wirelessly, facilitating improved quality and productivity in the industry.

In \cite{ref2} Ling et al. describes an XBee wireless protocol-based elderly infrared body temperature tele-monitoring system. The technology uses infrared body temperature sensors to remotely check on elderly people's body temperatures. The XBee protocol is used to wirelessly send the temperature data obtained, allowing for real-time monitoring and analysis. With the aid of this device, older patients' health can be monitored discreetly and conveniently, enabling early identification of aberrant temperature levels and fast intervention when required. However, detailed information regarding the affiliations of the authors or more data regarding the reference are not available.

In \cite{ref3} Sean et al. examines the idea of ambient assisted living (AAL) and how the Internet of Things (IoT) is integrated into it. For the purpose of creating an intelligent environment for assisted living, the authors suggest a system that makes use of a range of sensors combined with the XBee platform. These sensors keep an eye on a number of variables, including temperature, humidity, light, motion, and gas concentrations, and they provide real-time data for assessing people's safety and well-being. With the help of the XBee platform, wireless connectivity and communication are made possible, allowing for smooth data transfer between the sensors and the main system. The system aims to improve people's quality of life by offering a smart living environment that fosters independence, security, and comfort, especially for the elderly or people with disabilities.

In \cite{ref4} Allahham et al. provides a clever monitoring system for university settings that makes use of Zigbee WSNs. The system incorporates a network of wireless sensors placed throughout the campus to gather information on the climate, security, and energy use. These sensors enable real-time monitoring and analysis by wirelessly transmitting the data to a centralised monitoring system. The solution offers effective and dependable wireless communication while minimising power consumption by utilising Zigbee technology. The suggested system intends to better campus administration, boost safety precautions, maximise energy use, and offer useful information for making decisions. However, the reference does not provide any detailed information regarding the affiliations of the writers.

In \cite{ref5} Desnanjaya et al. presents a performance evaluation of the XBee Pro Series 2B RF module-based WSN for data transmission. The study focuses on measuring the efficiency and reliability of data transmission in a WSN context. The XBee Pro Series 2B RF module will be used in tests to monitor important performance parameters like packet loss, throughput, and latency. They examine how several elements, including distance, hop count, and interference, affect the effectiveness of the WSN. The analysis's findings shed light on the XBee Pro Series 2B RF module's advantages and disadvantages in terms of data transmission in a WSN configuration. 

In \cite{inproceedings} Maneesha V. Ramesh et al., and \cite{9034298} M Shyama et al.  the hindrance faced by WSN's were explored which gave us their significance of  these factors affect a WSN network.

In \cite{9848933} Sanjeev Kumar Shah et al., the study gave insights about the flexible and extensible architecture to integrate WSN and IoT, and in \cite{SANJAYKUMAR2020103275} V.Sanjay Kumar et al., the authors describe the effectiveness of sensory networks in safeguarding humans and were able to implement a smart surveillance system. 

\section{Methodology}

Hardware elements such as the Jetson Nano, Arduino microcontrollers, XBee modules, and other sensors are combined to build a smart home application. The Arduino microcontrollers operate as bridges between the Jetson Nano and the sensor nodes, while the Jetson Nano serves as the main controller. The Jetson Nano, Arduino microcontrollers, and sensor nodes can all communicate wirelessly with each other using XBee modules, which also makes it easier to send and receive control signals and data. The system's sensors gather information on variables related to the environment, including vibration, soil moisture, water levels, human presence, distance, temperature, humidity, flame, gas concentration, light intensity, sound intensity, and level of light and sound. Within the smart home system, LED lights, buzzers, and a 7-color LED provide visual and aural feedback.The collected data can be processed and analyzed by the Jetson Nano for decision-making and control of the connected devices.

The hardware setup includes the following components:

\begin{itemize}
    \item \textbf{Jetson Nano}: The Jetson Nano serves as the main controller and interface for the smart home system. It provides powerful computing capabilities and acts as the central hub for data processing and decision-making.
    \item \textbf{Arduino}: Four Arduino microcontrollers are utilized in the project. These microcontrollers serve as intermediaries between the Jetson Nano and the different sensor nodes, facilitating data acquisition and communication.
    \item \textbf{Breadboard}: Four breadboards are used to provide a platform for connecting and prototyping the various hardware components, enabling their integration into the system.
    \item \textbf{Jumper Wires}: Jumper wires are employed to establish electrical connections between different components, ensuring the proper flow of data and signals within the system.
    \item \textbf{XBee Modules}: Three XBee modules are utilized to enable wireless communication between the Jetson Nano, Arduino microcontrollers, and the sensor nodes. These modules facilitate the transmission of control signals and data exchange within the smart home application.
    \item \textbf{Sensors}: A variety of sensors are incorporated into the system to gather data on different environmental parameters. These include:
    \begin{itemize}
        \item \textbf{Flame sensor}: Detects the presence of fire or flames.
        \item \textbf{Gas sensor}: Measures the concentration of gases in the environment.
        \item \textbf{Photoresistor sensor}: Detects and measures light levels.
        \item \textbf{Big sound sensor}: Captures and analyzes sound intensity.
        \item \textbf{Temperature and humidity sensor}: Measures temperature and humidity levels.
        \item \textbf{PIR sensor}: Detects human presence based on infrared radiation.
        \item \textbf{Ultrasonic sensor}: Measures distance by emitting and receiving ultrasonic waves.
        \item \textbf{Shock sensor}: Detects sudden vibrations or movements.
        \item \textbf{Soil and moisture sensor}: Measures moisture levels in the soil.
        \item \textbf{Water level sensor}: Monitors the water level in tanks or containers.
    \end{itemize}
    \item Additional Components: The setup also includes LED lights, buzzers, and a 7-color LED for visual and auditory feedback.
\end{itemize}

\begin{table*}[!ht]
  \caption{\textbf{Comparative Discussion}}
  \label{tab:cd}
  \begin{tabular}{|p{2cm}|p{3cm}|c|c|c|}
    \hline
    \textbf{Feature} & \textbf{Zigbee} & \textbf{WiFi} & \textbf{LoRa} & \textbf{Bluetooth}\\
    \hline \hline
    Focus & Home Automation & High-Speed data & Low-power long range & Short range communication\\
    \hline
    Battery Life & Longer battery life & Shorter battery life & Very long battery life & Shorter battery life\\
    \hline
    Range & 100 m & 100 m & Several km & 10m \\
    \hline
    Bandwidth & Max 250 kbps & Upto several Gbps & Max 27 kbps & Max 24 kbps \\
    \hline
    Advantages & Low power mesh networks & High data rate , easy setup & Long range, low power & Ease of use, low cost, compatability.\\
    \hline
    Frequency Band & 2.4GHz, 915MHz, 868MHz & 2.4GHz, 5GHz & 433MHz, 868MHz, 915MHz & 2.4GHz\\
    \hline
    Maximum Nodes & 65,000 & 200 - 300 & Upto Millions & 7\\
    \hline
    Security & AES128 encryption & WPA/WPA2 encryption & AES encryption, spread spectrum & AES128 encryption\\
    \hline 
    Latency & 15 ms - 100 ms & A few ms to several 100 ms & 200 - 4000 ms & 1 ms - 10 ms \\
    \hline
    Scalability & Highly Scalable & Scalable & Highly Scalable & Moderate Scalable \\
    \hline
    Ease Of Intergration & Moderate & Easy & Moderate & Easy \\
    \hline
    Cost & Moderate & Moderate to high & Low & Low to moderate \\
    \hline
    Mobility & No & Yes & No & Yes \\
    \hline
    Data Rate & Variable,  depends on network size and configuration & Upto several Gbps & Upto 50 Gpbs & Upto 2 Mbps \\
    \hline
    Frequency Hoping & Yes & No & Yes & No \\
    \hline 
    Interference Tolerance & Good & Poor & Good & Poor \\
    \hline
    Device Complexity & Low & High & Low & Low \\
    \hline
    Bit error rate & $10^-6$ - $10^-9$ & $10^-6$ - $10^-9$ & $10^-5$ - $10$ & NIL \\
    \hline
    \end{tabular}
\end{table*}

\subsection{Xbee and its Role in WSN}
XBee is a wireless communication module that uses the Zigbee protocol. Zigbee is a low-power, low-range wireless protocol suitable for home automation. It operates in the 2.4 GHz ISM band and provides low power consumption and high security. XBee modules can be used for wireless communication, creating networks, and remotely controlling devices. In a wireless sensor network (WSN), XBee modules connect sensor nodes to a central controller like Raspberry Pi or Jetson Nano. They enable data transmission and remote control of devices within the smart home. XBee modules can also be configured as a mesh network, ensuring robust and fault-tolerant communication. XBee S2C documentation and datasheet is available in the link given here (\url{https://www.digi.com/resources/library/data-sheets/ds_xbee-s2c-802-15-4}).
\begin{figure}[htbp]
    \centering
    \includegraphics[width=0.3\textwidth]{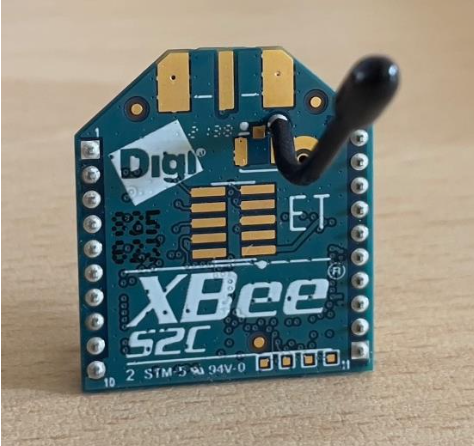}
    \caption{XBee Module}
    \label{fig:label_name}
\end{figure}

The methodology involves connecting and configuring these hardware components according to the desired system architecture. The Arduino microcontrollers interface with the sensors and XBee modules to collect sensor data and transmit it to the Jetson Nano. The Jetson Nano then processes and analyzes the data, utilizing machine learning algorithms if necessary, to make decisions and control the connected devices within the smart home.

The XBee modules enable wireless communication between the Jetson Nano and the Arduino microcontrollers, as well as between the Arduino microcontrollers and the sensor nodes. This wireless communication ensures seamless data exchange and control signal transmission throughout the smart home system.

It is important to note that, unlike traditional datasets, the data in this project is collected in real time from the interconnected Zigbee modules. The collected data is then utilized for analysis and decision-making within the smart home application.

This  methodology involves setting up the hardware components, configuring the Arduino microcontrollers and XBee modules, and establishing wireless communication between them. The sensors are integrated into the system to collect real-time data, which is processed by the Jetson Nano for decision-making and control. This hardware-based approach enables the development of a functional smart home application with the capability to monitor and control various aspects of the home environment.

\subsection{Setup Description}
In this section, we describe the implementation of our work. 
The proposed project involves using two Xbee modules to transmit data wirelessly from one location to another. The Xbee modules will be connected to multiple sensors and Arduino microcontrollers, which will collect data from the sensors and send it to the Xbee modules for transmission. The Image \ref{fig:SA} the architecture of a smart home system in which Bee is connected to sensors in every room. These sensors collect data on temperature, humidity, air quality, and other environmental factors.

\begin{figure}[htbp]
    \centering
    \includegraphics[width=0.5\textwidth]{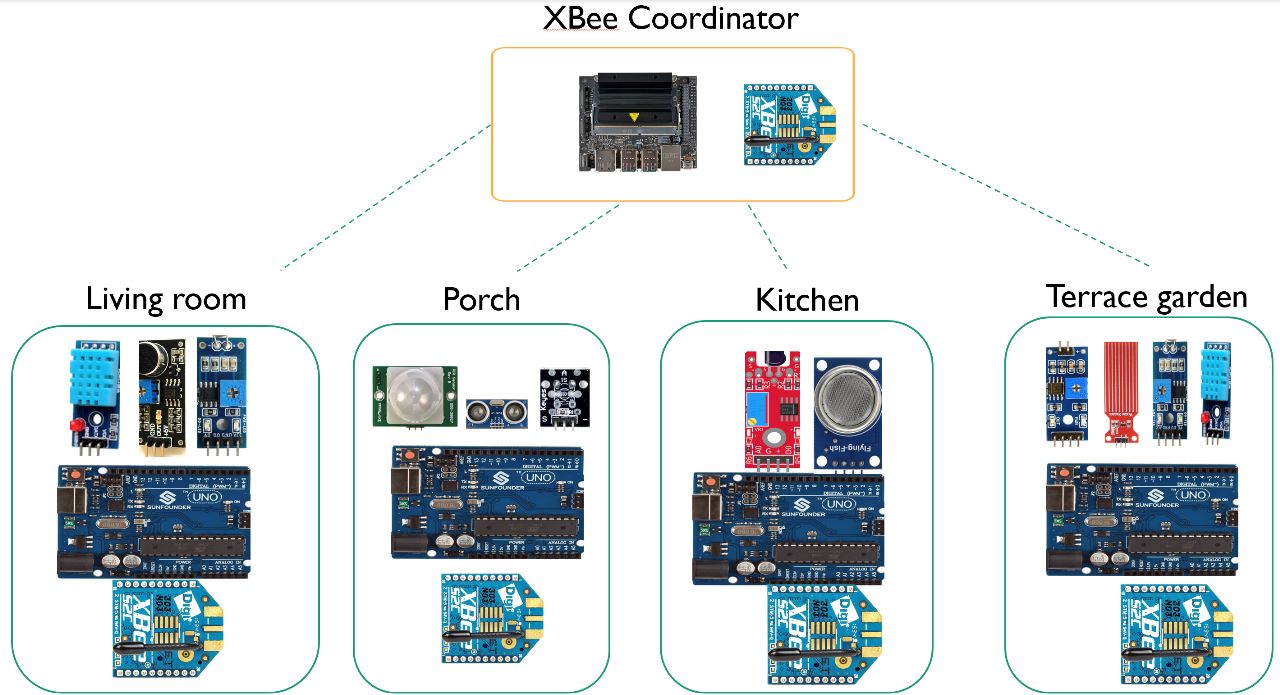}
    \caption{Setup Architecture}
    \label{fig:SA}
\end{figure}

We use a library called software serial library in our project enables us to use any digital pin on the micro controller as an RX or TX pin, which provides flexibility in terms of pin usage and allows us to communicate with other devices using serial communication without being limited to the 
hardware serial pins. Additionally, the library provides a number of useful functions for reading and writing data to the software serial port, making it easy to implement serial communication in our project.

\begin{itemize}

\item \textbf{Living Room} : a sound sensor to measure sound levels, a photoresistor to measure ambient light levels, and two pins for a disco light effect. The sound level is read using the sound sensor, the photoresistor is read to measure ambient light levels, and the DHT sensor is used to measure temperature and humidity. The function takes a string as input and sends it over the serial connection by breaking it down into bytes and sending each byte one at a time.The image depicts the sensors used in this room \ref{fig:lr}.

\begin{figure}[htbp]
    \centering
    \includegraphics[width=0.4\textwidth]{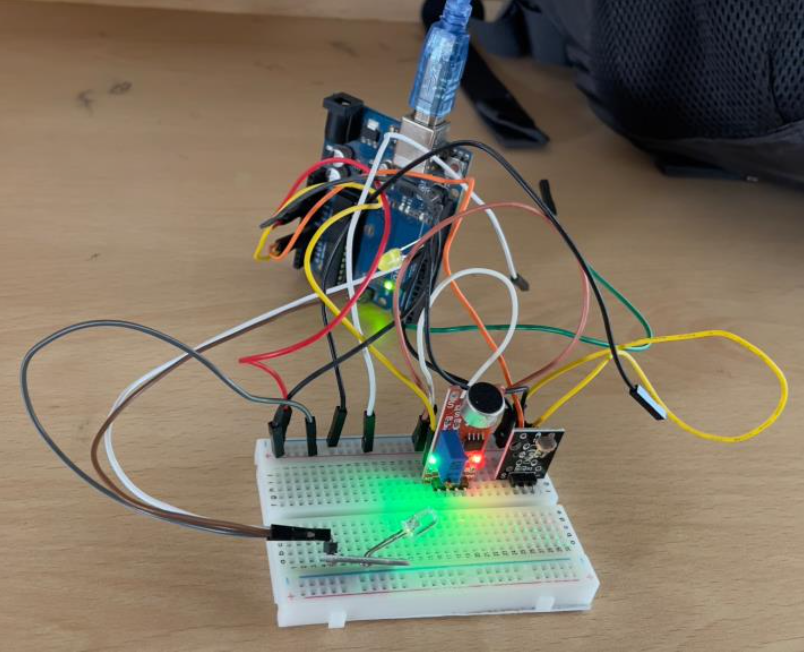}
    \caption{Living Room }
    \label{fig:lr}
\end{figure}

If the ambient light level is greater than 500, the disco light effect is activated and the two pins are set to opposite states. if the sound level is greater than 30, the led pin is turned on for 2 seconds. Finally, the loop function waits for 1 second before repeating.

\item \textbf{Kitchen} : uses two analog sensors, a flame sensor and a gas sensor, and a buzzer for an auditory output. The flame sensor and gas sensor are used to measure the values of flame and gas in the surrounding. If the flame sensor value is less than 800 or the gas sensor value is greater than 600, the buzzer sounds for 1 second. Finally, the loop function waits for 1 second before repeating.The image depicts the sensors used in this room \ref{fig:Kt}.

\begin{figure}[htbp]
    \centering
    \includegraphics[width=0.4\textwidth]{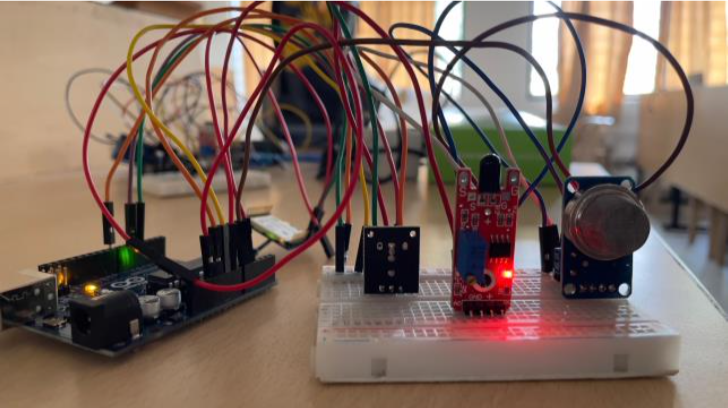}
    \caption{Kitchen}
    \label{fig:Kt}
\end{figure}

\item  \textbf{Porch} : Sensors used are ultrasonic sensor to measure distance, a PIR sensor to detect motion, and a shock sensor to detect shocks. It also uses a LED and a buzzer for visual and auditory output. The distance is measured using the ultrasonic sensor, the PIR sensor is read to detect motion, and the shock sensor is utilized to detect shocks. If a shock is detected, the LED is turned on. The distance, motion detection, and shock-sensor state are then printed to the Serial console and sent over the serial connection to another device using the "send message" function. The function takes a string as input and sends it over the serial connection by breaking it down into bytes and sending each byte one at a time. The image depicts the sensors used in this room \ref{fig:p}.

\begin{figure}[htbp]
    \centering
    \includegraphics[width=0.4\textwidth]{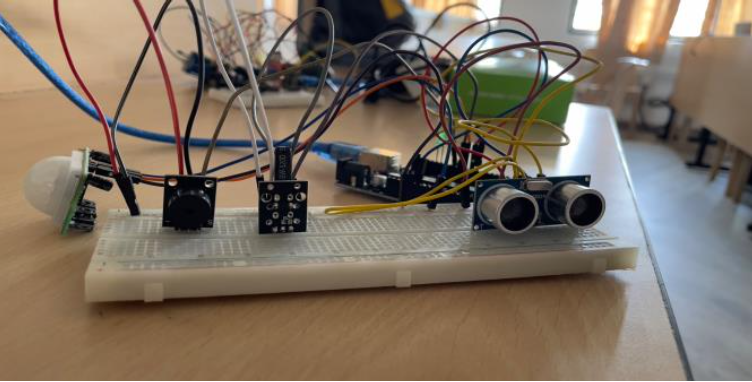}
    \caption{Porch}
    \label{fig:p}
\end{figure}

\item \textbf{Terrace garden} : We use a DHT11 sensor to measure temperature and humidity, an analog sensor to measure the moisture level of the soil, and another sensor to measure the water level in a tank. It also uses a buzzer for an auditory output. The moisture sensor and water level sensor are read to measure the values of moisture and water level. The DHT sensor is used to measure the temperature and humidity. If the water level sensor value is greater than 600, the buzzer sounds for 1 second. Finally, the loop function waits for 1 second before repeating. The image depicts the sensors used in this room \ref{fig:Tg}.

\begin{figure}[htbp]
    \centering
    \includegraphics[width=0.4\textwidth]{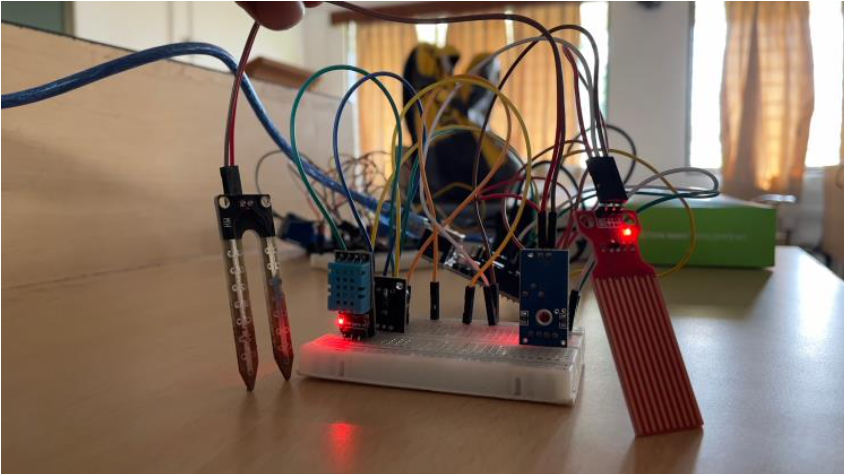}
    \caption{Terrace Garden}
    \label{fig:Tg}
\end{figure}

\end{itemize}

The data will then be received by the second Xbee module, which can be connected to another Jetson Nano for further processing or storage (MongoDB Cluster).

Overall, the system will allow for wireless monitoring and data collection from multiple sensors in different locations using Xbee wireless communication technology and Arduino microcontroller 
boards. 

\subsection{Cloud Integration}
In this scenario, a Wireless Sensor Network (WSN) is being set up using XBee radios, a Jetson 
Nano single-board computer, and MongoDB Atlas. The XBee radios are used to wirelessly 
transmit sensor data from the sensor nodes to the Jetson Nano, which acts as a gateway to the 
MongoDB Atlas database.MongoDB Atlas is a cloud-based, fully managed version of MongoDB, 
which is a popular NoSQL database. The sensor data collected by the XBee radios are stored in 
MongoDB Atlas, where it can be easily queried, analyzed, and visualized.

The Jetson Nano is programmed to communicate with the XBee radios and collect sensor data. It 
also uses the PyMongo library to interact with MongoDB Atlas. The Jetson Nano can be 
configured to establish a connection with the MongoDB Atlas server and insert the sensor data 
into a specific database and collection.
The sensor data stored in MongoDB Atlas can be accessed and visualized using various tools such 
as MongoDB Compass, a graphical user interface for MongoDB, or by using the MongoDB query 
language (MQL) to create custom queries and aggregations.

\subsection{GUI}
Streamlit is a Python library that allows you to create interactive web apps for machine learning 
and data science. To deploy a Streamlit app that uses data collected from Xbee transmitted to 
Jetson board and stored in MongoDB Atlas,the following steps are to be followed.
The app starts by connecting to the MongoDB database using the pymongo library and the 
MongoDB connection string. It then presents the user with a dropdown menu to select a room to 
analyze, which sets the collection to retrieve data from House.py .
The app then presents the user with another dropdown menu to select a field to analyze. It then 
displays the current time when the data was retrieved, and a line chart of the time series data for 
the selected field using the st.line-chart() function.
Next, we define a function get-data() that retrieves data from the selected collection, converts the 
timestamp field to a datetime object, sets it as the index and returns the current time and dataframe. 
This function is called to initialize the data when the app starts.
We create a Streamlit app that connects to a MongoDB database, allows the user to 
select a room and field to analyze, retrieves and displays the data in a line chart, and provides the 
ability to refresh the data.

\section{Results and Discussions}

We have successfully combined Jetson Nano, Arduino microcontrollers, XBee modules, and numerous sensors. The main controller and data processing component is a Jetson Nano, while the bridges to the sensor nodes are Arduino microcontrollers. Wireless connection is made possible via XBee modules, making it easier to exchange control signals and send and receive data. A wide variety of sensors are used in the system to collect data on environmental factors, and LED lights, buzzers, and a 7-color LED are used to provide visual and acoustic feedback. Real-time monitoring and control are made possible by the Jetson Nano's processing and analysis of the sensor data. In order to link sensor nodes to the Jetson Nano and enable remote device control, XBee modules create a wireless sensor network. MongoDB Atlas is used to store and manage sensor data, making it simple to query, analyse, and visualise. MongoDB is accessed through a graphical user interface (GUI) created with Streamlit, which offers an interactive platform for real-time analysis and visualisation of the sensor data. Overall, using this methodology allows for the creation of a smart home system that is functional and has monitoring, controlling, and analytical capabilities.

Table \ref{tab:cd} discusses that zigbee is suitable for home automation with low-power mesh networks and a maximum of 65,000 nodes. WiFi offers high-speed data transmission and easy setup, but with shorter battery life and limited range. LoRa excels in low-power, long-range communication, while Bluetooth provides short-range communication, ease of use, and compatibility. The choice of wireless technology depends on factors such as range, power consumption, data rate, security, and scalability.

\section{Conclusion and Futureworks}

We have successfully integrated Jetson Nano, Arduino microcontrollers, XBee modules, and various sensors to create a comprehensive smart home system. It enables real-time monitoring, control, and analysis of environmental variables through wireless communication, data processing, and cloud integration.

Potential  future works might include implementing a secured sytem as Zigbee networks are vulnerable to various types of cyber-attacks, so future work  in this area will focus on developing more secure Zigbee protocols and systems.
Edge computing: Zigbee networks can be integrated with edge computing architectures to 
enable data processing and decision-making at the network edge, which can lead to low latency and more efficient use of network resources.
The data collected can be used for implementing Machine learning based prediction
 Big Data Analytics: Zigbee networks generate large amounts of data, future work will 
focus on developing efficient algorithms to process, analyze and extract useful information 
from these data

\bibliographystyle{IEEEtran}
\bibliography{IEEEabrv, refs,aref1,aref2,aref3,aref4}

\end{document}